%% file: bare_conf.tex
\documentclass[conference]{IEEEtran}
\usepackage{cite}
\input{custom}

\begin{document}

\title{A PGAS Communication Library for Heterogeneous Clusters}

\author{
\IEEEauthorblockN{Varun Sharma}
\IEEEauthorblockA{Dept. of Electrical and\\Computer Engineering\\
University of Toronto\\
Toronto, Ontario\\
Email: varuns.sharma@mail.utoronto.ca}
\and
\IEEEauthorblockN{Paul Chow}
\IEEEauthorblockA{Dept. of Electrical and\\Computer Engineering\\
University of Toronto\\
Toronto, Ontario\\
Email: pc@eecg.toronto.edu}
}

\maketitle

\begin{abstract}
This work presents a heterogeneous communication library for clusters of processors and FPGAs. This library, \shoal{}, supports the Partitioned Global Address Space (PGAS) memory model for applications. PGAS is a shared memory model for clusters that creates a distinction between local and remote memory access. Through \shoal{} and its common application programming interface for hardware and software, applications can be more freely migrated to the optimal platform and deployed onto dynamic cluster topologies.

The library is tested using a thorough suite of microbenchmarks to establish latency and throughput performance.
We also show an implementation of the Jacobi iterative method that demonstrates the ease with which applications can be moved between platforms to yield faster run times.
Through this work, we have demonstrated the feasibility of using a PGAS programming model for multi-node heterogeneous platforms.
\end{abstract}

\IEEEpeerreviewmaketitle

\input{sections/introduction}
\input{sections/background}
\input{sections/shoal}
\input{sections/evaluation}
\input{sections/conclusion}

\section*{Acknowledgment}

The authors would like to thank Xilinx, CMC Microsystems, and NSERC for supporting and funding our research. 

\bibliographystyle{IEEEtran}
\bibliography{sample-base, references}

\end{document}

%% file: custom.tex
\usepackage[pdftex]{graphicx}
\usepackage{booktabs}
\usepackage{url}

\newcommand{\citeTip}[2][]{%
  \cite{#2}%
}

\newcommand{\shoal}{Shoal}

\graphicspath{{./images/}}

\def\cpp{{C\nolinebreak[4]\hspace{-.05em}\raisebox{.4ex}{\tiny\bf ++}}}

\usepackage{bytefield}
\usepackage{listings}
\usepackage[frozencache=true,cachedir=minted-cache]{minted} 
\usepackage{subcaption}
\usepackage{siunitx} 

\newcommand{\fakesixtyfourbits}[1]{%
\tiny
\count32=#1
\advance\count32 by #1
\the\count32%
}

\newcolumntype{H}{>{\setbox0=\hbox\bgroup}c<{\egroup}@{}}

\newlength{\tableindentlen}
\newcommand{\tableindent}[1][1]{$\hspace{#1\tableindentlen}$}

%% file: sections/introduction.tex
\section{Introduction}
\label{ch:intro}

Heterogeneous computing through the incorporation of FPGAs in the data center has yielded beneficial results \cite{putnam_reconfigurable_2014, caulfield_cloud-scale_2016, firestone_azure_2018}.
Users who want to add FPGAs into their application can now do so through offerings from cloud vendors such as Amazon \cite{amazon_amazon_2020} and Nimbix \cite{nimbix_nimbix_online} or implement their own custom cluster.
However, incorporating FPGAs into applications can be challenging.
The standard tool flows around FPGAs require experienced digital hardware experts and result in long development times when compared to software.
Furthermore, integrating these devices into heterogeneous platforms requires developing a communication scheme that can work across both processors and FPGAs as well as providing the low-level infrastructure for basic I/O.
The familiar challenges of using FPGAs magnify at scale and result in the inability to easily leverage the available resources.

Our work focuses on a major difficulty facing applications in such a system: effective communication between devices.
The chosen method of communication must be convenient, easy to use, scalable, and adaptable for different scenarios.
Prior work has identified the PGAS programming model as a good fit to meet these requirements \cite{shan_preliminary_2012} and developed a heterogeneous communication infrastructure~\cite{willenberg_heterogeneous_2014}.
Due to limitations of the implementation, the infrastructure developed in~\cite{willenberg_heterogeneous_2014} cannot be used at scale in a data center environment.
In our work, we refine past implementations and combine them with more recent developments in managing heterogeneous clusters to more fully address the problem of communication in the data center.

Our major contributions are:
\begin{itemize}
    \item A new heterogeneous PGAS communication library, \shoal{}, and the associated hardware IPs to facilitate communication between hardware and software.
    \item Hardware and software infrastructure that is compatible with Galapagos~\citeTip{eskandari_modular_2019}, which is an open-source framework that provides deployment, connectivity between nodes, and enables usage of the library in dynamic clusters.
    \item A characterization of the \shoal{} API through microbenchmarks.
    \item An adaptation of the Jacobi method application from prior work to demonstrate the functionality of \shoal{} in a realistic use case.
\end{itemize}

The remainder of this paper is organized as follows.
Section \ref{ch:background} presents basic information on memory models and Galapagos~\citeTip{eskandari_modular_2019} and some prior work.
Section \ref{ch:shoal} describes the main body of this work: the \shoal{} platform and its API.
Section \ref{ch:evaluation} evaluates the \shoal{} platform through microbenchmarks and then presents experimental results from the Jacobi application adapted from previous work.
Section \ref{ch:conclusion} concludes the paper and examines future work.

%% file: sections/background.tex
\section{Background}
\label{ch:background}

In this section, we first compare three memory models used in programming.
In particular, the Partitioned Global Address Space (PGAS) model used in this work is defined here.
Next, we present Galapagos, a platform for creating and deploying heterogeneous clusters.
The use of Galapagos in this work greatly simplifies many of the low-level details that would otherwise be developed from scratch.
Finally, we summarize some related work in the area of PGAS programming and communication libraries.

\subsection{Memory Models}
\label{sec:background:memory}

\begin{figure}[ht]
  \centering
  \includegraphics[width=\linewidth, page=4]{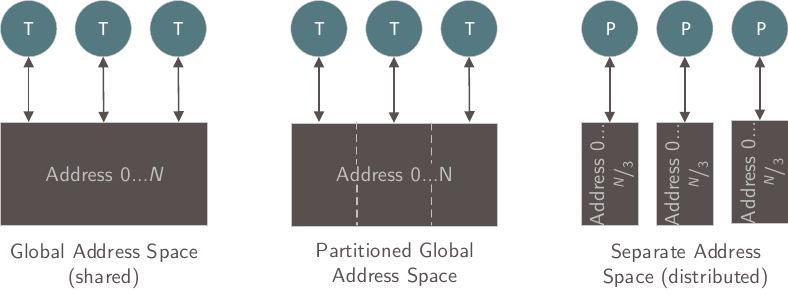}
  \caption[Memory models with different threads and processes.]{Memory models with different threads and processes (adapted from~\citeTip{willenberg_heterogeneous_2014})}
  \label{img:background:memory}
\end{figure}

A memory model describes the developer's abstract view of system memory for parallel computations.
There are three models that we describe here: shared, distributed and the partitioned global address space (PGAS), illustrated in Figure~\ref{img:background:memory}.
In this section, we use the term ``node'' to refer to one parallel computing element such as a thread or a process.

\subsubsection{Shared}
\label{sec:background:memory:shared}

In a shared memory model, all parallel parts of an application have access to the same memory, which can be accessed using the same addresses.
Data allocated in one parallel compute element are accessible in others without explicit communication.
Multithreaded programs inherently use shared memory between different threads.

The advantage of using a shared memory model is its ease of use.
There is no need for explicit communication between different nodes since all nodes can access all data.
Unfortunately, the shared memory model does not scale well across many machines due to the expense associated with creating shared memory.

\subsubsection{Distributed}
\label{sec:background:memory:distributed}

The distributed memory model is the traditional one used in multi-node high performance computing clusters.
Each node in the cluster has its own separate memory address space.
To share data between nodes, explicit communication between them must take place.

The message passing paradigm is frequently used in this space to facilitate nodes exchanging data.
The \textit{de facto} implementation of this protocol is the aptly named Message Passing Interface (MPI)~\citeTip{clarke_mpi_1994}.
While the MPI specification has grown to include more exotic features such as one-sided remote memory access~\citeTip{geist_mpi-2_1996}, the core of the protocol is as follows.
Communication is between two nodes and considered \textit{two-sided}, which is to say that both nodes must participate in the exchange.
The simplest use of MPI is where one node sends data while the other node blocks execution until it receives the message.
Two-sided communication also forces the communicating parties to stop potential useful work, perform handshaking and wait for the data transfer.

\subsubsection{Partitioned Global Address Space}
\label{sec:background:memory:pgas}

The PGAS model is a hybrid of the shared and distributed models.
The memory of different nodes are physically separate as in the distributed model but logically contiguous in that any node can access the shared space~\citeTip{almasi_pgas_2011}.
Bridging the disparate memories together requires support from hardware or software.

PGAS attempts to combine the best of both worlds: to gain the benefits of easier application writing and to scale across many machines in a cluster.
Each node in a PGAS cluster has one partition of the global address space.
While all nodes can access the shared space, this locality information is known to the programmer and so accessing data stored in other partitions is implemented as a remote access.
PGAS promotes the concept of \textit{one-sided} communication in which only the application on the participating node is explicitly involved in the communication, which makes it easier to overlap communication and computation as compared to MPI~\citeTip{shan_preliminary_2012}.
As in all memory models, poorly optimized data locality in PGAS results in performance penalties.

\subsection{Galapagos}
\label{sec:background:galapagos}

Galapagos is an open-source stack and middleware platform to support the creation of heterogeneous clusters through a set of user-provided configuration files~\citeTip{tarafdar_galapagos_2018}.
As our work is built on top of and relies on Galapagos, we provide a description of this framework and explain why it is a good fit for this project.

Galapagos is the source of the definitions of ``node'' and ``kernel'' we use in the remainder of this work.
A node, unless otherwise stated, refers to a processor, FPGA or another device in a cluster that has a unique network address such as an IP address.
Each node may have one or more kernels on it, where each kernel is an independent computing element (for example a thread or a hardware IP) that has a globally unique kernel ID assigned to it.

\subsubsection{A Layered Approach}
\label{sec:background:galapagos:layers}

\begin{figure}[ht]
  \centering
  \includegraphics[width=\linewidth, page=3]{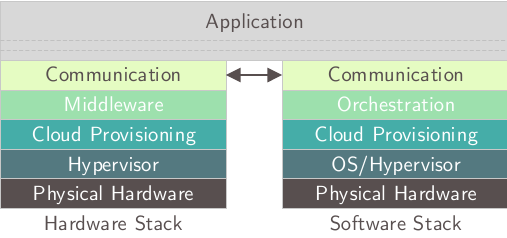}
  \caption[The Galapagos Stack]{The Galapagos stack (adapted from~\citeTip{tarafdar_galapagos_2018})}
  \label{img:background:gal_stack}
\end{figure}

One of the central tenets of Galapagos is the importance of layers.
This idea is not original to Galapagos.
In the OSI model, different layers of the network stack hide implementation-specific details of lower layers through interfaces~\citeTip{briscoe_understanding_2000}.
Different protocols may be used within a layer provided they present the same interfaces to the layers above and below.
The layers of Galapagos serve the same purpose and make hardware and software nodes more similar to each other.
Through heterogeneous communication, distributed applications can exist on either hardware or software transparently.
The layers of the heterogeneous Galapagos stack are shown in Figure~\ref{img:background:gal_stack} and are discussed in more detail in~\citeTip{tarafdar_galapagos_2018}.

\subsubsection{Why Galapagos?}
\label{sec:background:galapagos:why}

The notable advantage that Galapagos provides is portability across networks and simplified access to FPGA resources.

Through its dynamic cluster creation, Galapagos supports scalability over the network where FPGAs can exist as first-class computation nodes.
The incorporation of layers also allows transparent changes under the hood with no user impact.
For example, Galapagos currently supports TCP, UDP and raw Ethernet packets for communication---which can be chosen in the Middleware layer and changed transparently to the application---and determines which hardware cores are instantiated in the design or which protocol is used in software.
New protocols can be added and leveraged with little change in user code.

Many of Galapagos' functionalities are critical for any heterogeneous cluster application and having them codified into a framework simplifies usage and encourages reuse rather than reinvention.
In hardware, access to the network and off-chip memory would have to be set up regardless but would otherwise be done manually.
Galapagos separates the FPGA into two regions: the Shell and the Application Region.
The reusable Shell hosts the static interfaces to external devices while the user designs reside in the Application Region and change more frequently.
In software, libGalapagos presents a stream-based protocol that can be used to send data between kernels instead of setting up the sockets or interacting with a network library directly.
Finally, routing data to the correct destination is essential for both hardware and software kernels in a cluster environment, which Galapagos manages instead of requiring the user to contrive a scheme.

\subsection{Related Work}
\label{sec:background:related}

This section reviews some prior work in this area.
First, Section~\ref{sec:background:gasnet} discusses GASNet: a communication standard that influenced a number of later efforts in PGAS development, including those that inspired this work.
Then, Section~\ref{sec:background:thegasnet} briefly summarizes two works, THeGASNets, that directly precede our work.
Finally, the HUMboldt protocol is presented in Section~\ref{sec:background:humboldt}.
This lightweight implementation of MPI was the first language abstraction built on top of Galapagos.

\subsubsection{GASNet}
\label{sec:background:gasnet}

GASNet~\citeTip{bonachea_gasnet_2017} is a specification for a low-level communication layer that can be used as part of higher-level PGAS languages.
The specification defines a language- and platform-independent interface for implementations to use.
GASNet implementations are used under languages like UPC~\citeTip{el-ghazawi_upc_2005}, Co-Array Fortran~\citeTip{numrich_co-array_1998}, and Chapel~\citeTip{chamberlain_parallel_2007} to seamlessly provide communication.

The Core API defines methods to initialize a GASNet program and for nodes\footnote{Node, in GASNet jargon, refers to a single compute element, whether that is a thread or a processor, that runs an instance of the application and therefore has an address space and an ID. This definition of node is used in the context of GASNet.} to query information about themselves and other nodes.
The Extended API defines some higher-level functions such as blocking and non-blocking get/put messages to exchange data.

Active Messages (AMs) are the mechanism through which nodes in GASNet communicate.
Originally proposed in~\citeTip{von_eicken_active_1992}, AMs differ from conventional messaging in that they can trigger computation upon receipt through the use of handler functions.
Handler functions are user-defined functions that may accept arguments and perform work.
Handler functions work as follows.
Node A includes a particular handler ID when sending a message to node B.
After the received message is managed appropriately, the handler function associated with the handler ID will be called on node B to take additional action.

GASNet bases the AM definitions in its Core API on a modified version of the Active Messages 2.0 specification~\citeTip{mainwaring_active_1995}.
There are three main classes of AMs---Short, Medium and Long---that differ in content: short messages carry no payload, medium messages carry payload to a temporary buffer, and long messages carry payload to shared memory.
After processing the handler function, the remote node responds with a reply message back to the source node.

\subsubsection{THeGASNets}
\label{sec:background:thegasnet}

THeGASNet~\citeTip{willenberg_heterogeneous_2014} and THe\_GASNet Extended~\citeTip{pandit_extended_2016} (collectively referred to as THeGASNets) are related frameworks providing runtime support for heterogeneous PGAS applications.
They are adaptations of the Core and Extended GASNet APIs, respectively.
While these implementations are not scalable and impose unnecessary constraints, they also present concepts, such as in-built strided memory access for kernels, that are interesting.
We refine some of the ideas explored in these old works and apply them to modern heterogeneous clusters at scale in our work.

\subsubsection{HUMboldt}
\label{sec:background:humboldt}

The HUMboldt protocol \citeTip{eskandari_modular_2019} is a minimal implementation of MPI built on Galapagos for heterogeneous communication.
In particular, the library defines two functions \texttt{HUM\_Send()} and \texttt{HUM\_Recv()} to send and receive data to/from kernels.
These functions are provided in a \cpp{} header file that can be used in both software and hardware kernels (the latter using HLS).
HUMboldt, as in MPI, uses two-sided communication where all participating kernels must be involved in the data exchange.
Communication is initiated by the sending a request that the receiver acknowledges.
At this point, the sender is cleared to send data.
The receiver sends a final message back to the sender to complete the transaction.

%% file: sections/shoal.tex
\section{\shoal{}}
\label{ch:shoal}

\shoal{} is our implemented heterogeneous PGAS library that defines an API for Active Message (AMs) communication between kernels as well as barriers for synchronization.
The AMs used in \shoal{} are based on the AMs used in THeGASNet.
It supports widely distributed workloads spread out over both hardware and software kernels over a generic network.
The network-agnosticism comes from the use of Galapagos and allows \shoal{} to take advantage of the abstraction provided.
In software, \shoal{} includes a \cpp{} library and headers for users to compile applications.
These applications can be run through an HLS tool or rewritten in Verilog  and moved to an FPGA.
Hardware support of the PGAS model is provided primarily through a hardware IP called the GAScore: a direct memory access (DMA) engine to facilitate remote memory access.

First, we describe the communication API used in \shoal{} in Section~\ref{sec:shoal:api}.
Sections~\ref{sec:shoal:sw} and~\ref{sec:shoal:hw} present the software and hardware implementations, respectively.
Finally, Section~\ref{sec:shoal:hw:galapagos} comments on how \shoal{} integrates with Galapagos.

\subsection{Communication API}
\label{sec:shoal:api}

The \shoal{} API is heterogeneous and uses the same (or very similar) function prototypes for software and hardware targets and has platform-specific function implementations where needed.
It defines three classes of AMs: Short, Medium and Long.
Short message types are primarily used for signaling and reply messages.
Medium message types serve as point-to-point communication for one kernel to send data directly to another kernel.
Finally, Long message types contain payload that is written to remote memory.
We carry forward support for Strided and Vectored Long messages as well.

The Medium and Long message types are further divided into two cases depending on the source of their payload data.
\textit{Medium/Long FIFO} messages are used to denote messages whose payload originates from the kernel while \textit{Medium/Long messages} specify messages whose payload comes from shared memory and is added by the runtime.
As described, these messages are the \textit{put} variants.

The API also supports \textit{get} requests for Medium and Long message types that can bring data from remote memory.
For example, a Medium \textit{get} request would bring remote data from a particular address in the destination kernel to the source kernel.

Each received packet triggers a reply unless the initial message is marked as asynchronous.
Reply messages are Short messages that trigger a handler function that increments a variable and thus keeps track of the number of reply messages received at each kernel.
Kernels can therefore send several messages and then collectively wait for the same number of replies to know that all have been received.

Runtime support for AMs is managed by the handler thread in software and by the GAScore in hardware, which are described in Sections~\ref{sec:shoal:sw} and \ref{sec:shoal:hw}, respectively.
These components are responsible for parsing incoming AMs and directing them appropriately, calling handler functions, and sending out AMs from local kernels.

In GASNet and THeGASNets, custom user-defined handler functions can be defined in software kernels.
While this functionality has been maintained in \shoal{} software kernels as well, it is not as applicable in hardware.
Supporting similar behavior in hardware requires a custom handler IP external to the other \shoal{} IPs that the user can modify as needed, or by adding a core internally that defines basic operations that a user can string together into the desired functionality.
In practice, this broad freedom is rarely needed and so it is removed to simplify the hardware implementation.
However, custom handler support can be added to hardware through existing workarounds or as future work.

In \shoal{}, management of reply messages has been absorbed into the runtime and managed without kernel intervention.
HLS can be used to simplify creating \shoal{} packets in the right format through the API.
In this way, a simple controller can be developed to send AMs based on custom control signals from the user IP.

\subsection{Software Implementation}
\label{sec:shoal:sw}

Software \shoal{} kernels are defined through a function pointer to a \textit{kernel function} that gets started as a separate thread by the software \shoal{} node.
Each kernel function has at least three arguments: an integer ID and a pair of Galapagos Interfaces (GIs) to send and receive data from other kernels.
GIs are defined in libGalapagos as stream-based heterogeneous interfaces that kernels use to communicate.
All local kernels on the node communicate using a router thread in libGalapagos while data for external kernels are routed from this router to an external driver such as TCP.

To add the PGAS memory semantics, we add a \texttt{handler thread} for each software kernel.
The handler thread is responsible for tasks such as parsing headers, redirecting data to memory or to kernels, and calling handler functions.
It serves as the gatekeeper between a particular kernel and the wider network.
Incoming messages may be redirected to shared memory or passed on to the kernel or trigger the appropriate handler function that an AM specifies.
Outgoing message requests from the kernel are read and converted into network packets to be sent out.

\subsection{Hardware Implementation}
\label{sec:shoal:hw}

The hardware implementation on \shoal{} relies primarily on the GAScore, which facilitates access to remote memory.
The GAScore is shared among all kernels on a node unlike handler threads that are created per kernel.

\begin{figure}[ht]
  \centering
  \includegraphics[width=\linewidth, page=5]{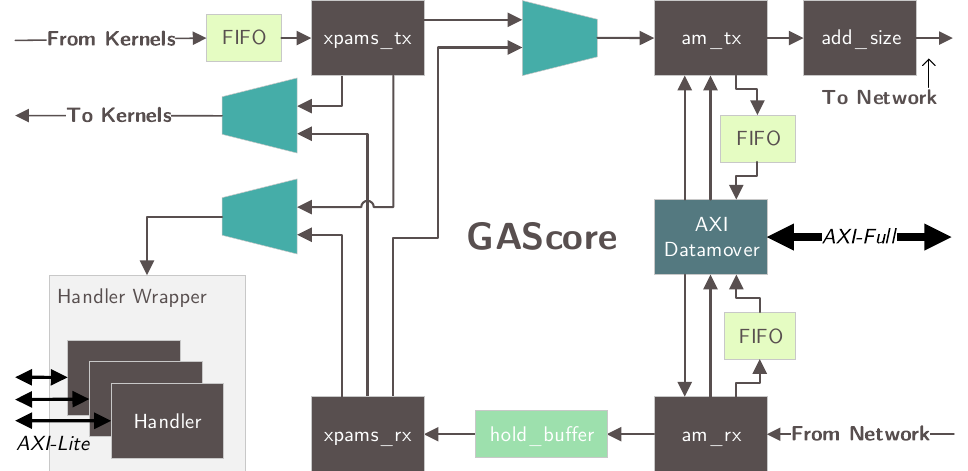}
  \caption{Internal structure of the GAScore in \shoal{}}
  \label{img:shoal:gascore}
\end{figure}

The GAScore is composed of IPs written through a mix of HLS and Verilog and its structure is shown in Figure~\ref{img:shoal:gascore}.
Unless otherwise indicated, the connections between these blocks are AXI4-Stream (AXIS) interfaces.
It has two pairs of master-slave AXIS interfaces: to/from the external network and to/from local kernels.
The GAScore provides a set of AXI-Lite slave interfaces---one for each local kernel on the FPGA---that are used to access the registers associated with the built-in handler functions.
Finally, the GAScore has one AXI-Full master interface to the shared memory.

The Xilinx AXI DataMover IP~\citeTip{xilinx_inc_axi_2017} is used to facilitate access to memory by providing a conversion between AXIS and AXI-Full.
A simplified explanation is provided here.
The DataMover exposes two parallel interfaces for reading and writing data.
To read, a read command is sent to the IP that specifies the source address and the number of bytes to read.
The read data is streamed back over AXIS and a status indicator is available.
Similarly, writing data to memory first requires a write command to the IP that specifies the destination address and the number of bytes to write.
After the command is sent, data can be streamed into the IP over AXIS and it is written to memory as instructed in the command. 
The \texttt{am\_rx} and \texttt{am\_tx} modules perform these actions when data must be written to or read from memory.

The general path of egress packets is described below, with some special cases excluded here for simplicity.
\begin{enumerate}
    \item A \shoal{} kernel packet arrives at the ``From Kernels'' interface.
    \item Based on the header, the type of message and destination is decoded in \texttt{xpams\_tx}. For the special cases of Short messages and Medium FIFO messages intended for local kernels, this module will route data to the handler internally and pass any payload directly to the ``To Kernels'' interface. Other messages types, whether they are to local or remote kernels, need access to memory and so proceed unaltered to \texttt{am\_tx}.
    \item \texttt{am\_tx} determines the type of message based on the header and parses the rest of the command packet. Depending on the message type, additional data may be added to the outgoing message. In particular, for messages with a payload, requests for data are sent over the DataMover's command interface and the read data from the IP is padded onto the end of the outgoing packet.
    \item The \texttt{add\_size} block adds the metadata needed for Galapagos to handle the data correctly. The final message size is counted and this size (in words) is added to the TUSER side channel of the AXIS interface.
\end{enumerate}
Ingress packets follow a similar path in the opposite direction.
\begin{enumerate}
    \item A \shoal{} packet arrives at the ``From Network'' interface.
    \item \texttt{am\_rx} parses the header and forwards it. For Long message types, the payload gets written to memory. The \texttt{hold\_buffer} is a special FIFO that buffers the forwarded data in the case of Long AMs. While the payload is being written to memory, the AM's header is held at the buffer. After it has been written, the message is allowed to proceed.
    \item Using the forwarded message, \texttt{xpams\_rx} will pass the handler function data on to the handlers. It will also forward the Medium AM payload to the kernels through the ``To Kernels'' interface. Finally, this module creates a reply packet and sends it to \texttt{am\_tx} to be sent back to the source kernel.
\end{enumerate}

\subsection{Integration with Galapagos}
\label{sec:shoal:hw:galapagos}

As described in Section~\ref{sec:background:galapagos}, Galapagos provides scripts to automate the creation of bitstreams.
We integrate with this flow in both the software and hardware implementations of \shoal{}.

In software, the \shoal{} API extends and calls libGalapagos to initialize nodes and communicate between kernels.

In hardware, we enforce that the ingress and egress packets to/from the GAScore are compatible with Galapagos.
We also modify the standard Galapagos script generation flow to add extra instructions to modify the user region to insert the GAScore and reconnect other IPs appropriately.
The hook that we have added into the Galapagos middleware for this purpose is generic and can be leveraged to modify the user region in a Galapagos FPGA in other ways.

%% file: sections/evaluation.tex
\section{Evaluation}
\label{ch:evaluation}

This section evaluates and characterizes the \shoal{} platform.
The software tests are run on servers with an Intel Xeon E5-2650 processor and 64GB of 2400MHz DDR4 RAM.
The hardware tests use one or more Alpha Data 8K5 boards, each of which has a Xilinx Kintex Ultrascale FPGA.
The servers and the FPGAs are connected to the same Dell S4048-ON 10G switch.
While these tests are run on this hardware configuration, there are no inherent restrictions preventing usage on other platforms.
This freedom is one of the advantages of Galapagos and so other boards and FPGAs can be used if corresponding Galapagos Shells exist.

First, Section~\ref{sec:eval:util} discusses the hardware utilization of Galapagos, the GAScore and the addition of \shoal{} API calls to HLS IPs.
Section~\ref{sec:eval:ubench} reports the performance of the API through microbenchmarks on a custom Benchmark IP.
Finally, an implementation of the Jacobi method is tested in Section~\ref{sec:eval:jacobi} as a real-world application.

\subsection{Hardware Usage}
\label{sec:eval:util}

On its own, using the \shoal{} API on an FPGA is relatively lightweight in terms of hardware resource usage.
Of course, its usage is predicated on the presence of some basic Galapagos infrastructure in the Shell and the Application Region.
However, as noted in~\citeTip{eskandari_modular_2019}, the use of Galapagos components does not necessarily contribute to higher utilization cost since they would have to be replaced with equivalent functional blocks in any system that does not use Galapagos.
Here, the Shell consumes about 12\%, 8\% and 8\% of the LUT, FF, and BRAM resources on the 8K5, respectively.
The off-chip memory controller and the PCIe controller dominate this usage.
Depending on the application, these components may be removed from the Shell if they are not needed to ease congestion and free resources.

\shoal{} hardware is found within the Application Region.
One GAScore IP must exist in this region and is shared among all local kernels.
The utilization of this IP is shown in Table~\ref{table:util:gascore:1} when there is one kernel present on the FPGA.
With more kernels, the \texttt{Handler Wrapper} grows approximately linearly in usage, and a handler is added for each kernel.
However, the additional cost of a larger interconnect between the different handlers grows as well.
The other subcomponents of the GAScore are shared between any additional kernels and remain constant in usage.
We see that the overhead to add PGAS capabilities on top of Galapagos is small---under 8000 LUTs and FFs and fewer than 30 BRAMs for one kernel---with each additional kernel consuming a few hundred more LUTs and FFs.
The total available resource counts for the FPGA on the 8K5 is shown in the last row for comparison.

\input{tables/utilization_GAScore_1}

\subsection{Microbenchmarks}
\label{sec:eval:ubench}

Measuring the results of microbenchmarks for the \shoal{} API requires sweeping many hardware configurations through many different types of communication methods.
Since Galapagos kernels may exist in both software and hardware, performance between all combinations of these placements must be measured.
There are six combinations: software-to-software (same node), software-to-software (different nodes), software-to-hardware, hardware-to-software, hardware-to-hardware (same node) and hardware-to-hardware (different nodes).
For packets with payloads, payload sizes from 8 bytes to 4096 bytes are tested.

The breadth of microbenchmarks demonstrates one of the advantages of using a platform like \shoal{} on a heterogeneous cluster.
With a single application source file in \cpp{}, we can run it on any platform in any topology, which lets developers focus on application writing rather than deployment as in the software world.
With FPGAs, we can dynamically change how an application gets mapped onto hardware and explore the design space for different data partitioning.

The tests described above are run to measure both latency and throughput.
For latency, time is measured from when the Sender sends the message to when it receives the reply from the Receiver.
For throughput, the Sender sends all the messages in a loop and then waits for all the replies.
We present the latency and throughput results in Sections \ref{sec:eval:ubench:shoal:latency} and \ref{sec:eval:ubench:shoal:throughput}, respectively.

\subsubsection{Latency}
\label{sec:eval:ubench:shoal:latency}

\begin{figure}[ht]
  \centering
  \includegraphics[width=\linewidth]{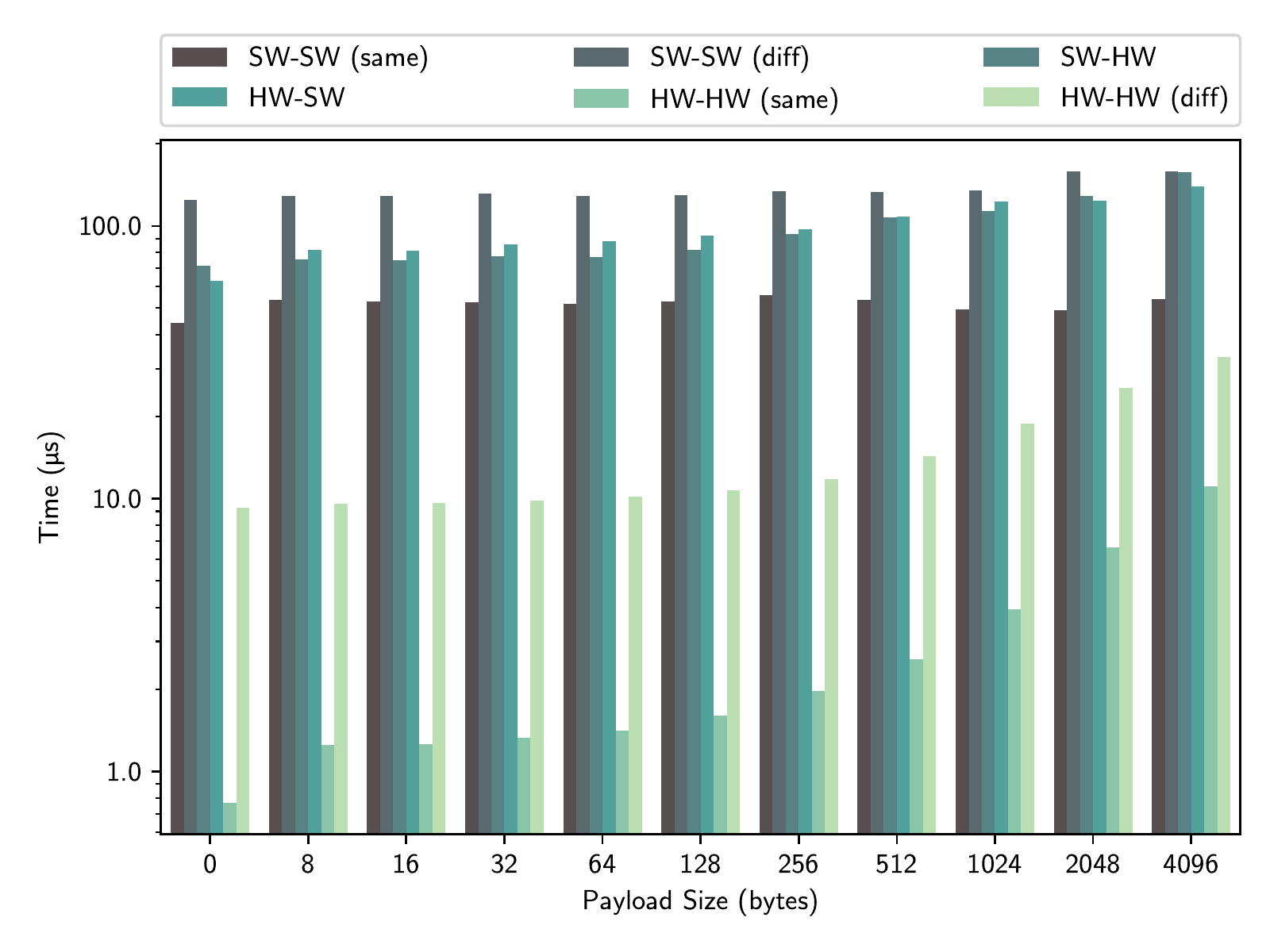}
  \caption{Average median latency of communication methods with TCP in different topologies}
  \label{img:eval:shoal:latency:tcp}
\end{figure}

The latency of communication is important to benchmark for a low-level communication API.
For the test cases described above, we perform the tests using both TCP and UDP, where possible.
Figure~\ref{img:eval:shoal:latency:tcp} shows the median latencies in different hardware topologies using TCP to communicate between kernels on different nodes.
Kernels on the same node use internal routing within the FPGA (in the hardware case) or within libGalapagos (in the software case).
For simplicity, this figure shows the average of the different types of AMs in each topology. 
As expected, communication between kernels in hardware occurs much faster than communication in software.
Even two hardware kernels on different nodes can use the whole TCP/IP stack faster than software can internally route data in libGalapagos.
For most cases, the latencies increase with increasing payload size.
Notably, SW-SW (same) shows a constant trend, indicating that there are other overheads beyond the payload size.

\begin{figure}[ht]
  \centering
  \includegraphics[width=\linewidth]{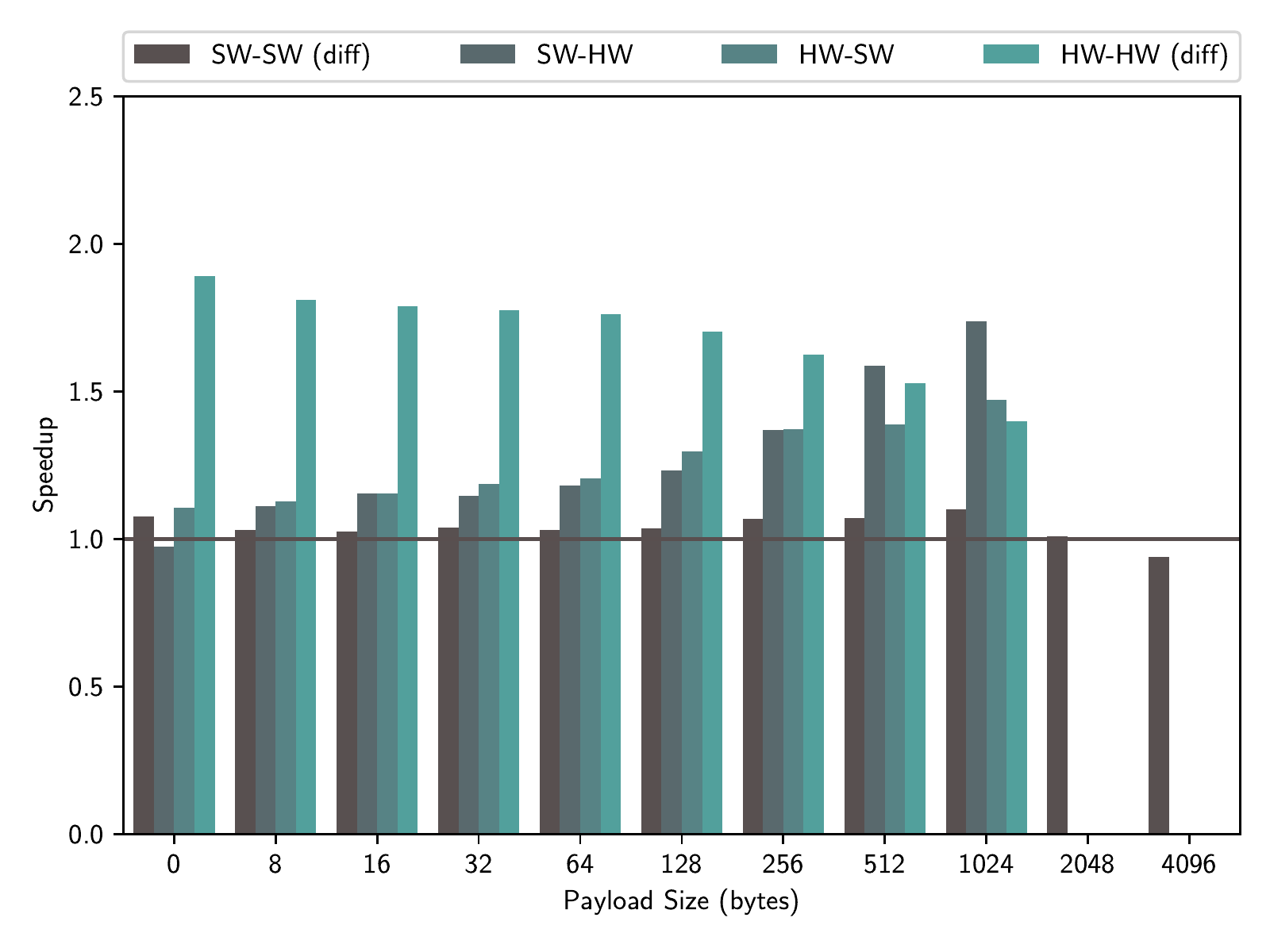}
  \caption{Speedup of median latency using UDP instead of TCP}
  \label{img:eval:shoal:latency:udp}
\end{figure}

Using UDP offers even shorter latencies.
Figure~\ref{img:eval:shoal:latency:udp} shows the speedup when using UDP instead.
This figure excludes the communication between kernels on the same node as no network protocol is used.
In most cases, messages sent with UDP are faster.
No data was collected for topologies including hardware for UDP messages with 2048 and 4096 byte payload sizes.
Large UDP packets sent from software are marked as IP fragmented, which is unsupported by the hardware UDP core on the FPGA.
The inverse case—transmitting large UDP packets from the hardware—fails as well because the UDP core does not send them out.
These packets may have been dropped by the core or are otherwise unsupported.
There are workarounds for this problem.
Data could be pre-fragmented to fit within the Ethernet frame and reconstructed using the Galapagos header.
However, this preprocessing has not been implemented.

As compared to the base Galapagos infrastructure, \shoal{} adds latency overhead primarily through packet parsing.
In addition, the GAScore is currently modular in design.
By more tightly integrating the different components, packet latency through it can be further reduced.

\subsubsection{Throughput}
\label{sec:eval:ubench:shoal:throughput}

\begin{figure}[ht]
  \begin{subfigure}{\linewidth}
    \centering
    \includegraphics[width=\linewidth]{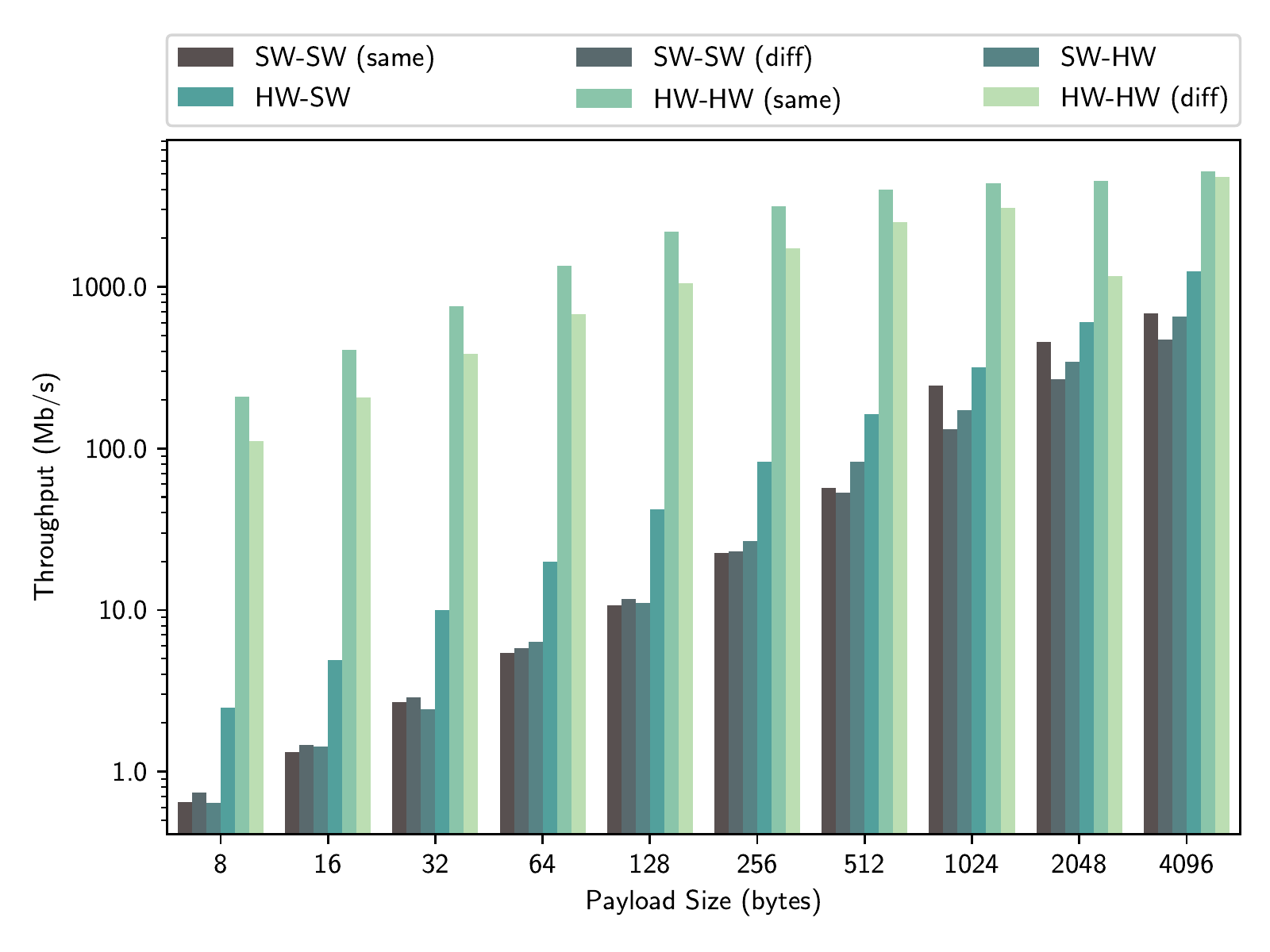}
  \end{subfigure}\hfill
  \caption{Average throughput of communication methods with TCP in different topologies}
  \label{img:eval:shoal:throughput:tcp}
\end{figure}

The throughput of communication measures the sustained data transmission rate.

Figure~\ref{img:eval:shoal:throughput:tcp} shows the throughput in different hardware topologies using TCP to communicate between kernels on different nodes.
As in the latency case above, kernels on the same node use internal routing within the FPGA (in the hardware case) or within libGalapagos (in the software case).
For simplicity, this figure shows the average of the different types of AMs in each topology.
Throughput between hardware nodes is significantly higher than software nodes and it generally increases with payload size.
With 4096 bytes of payload, the throughput between hardware kernels on different FPGAs in non-blocking communication is close to kernels on the same FPGA.

While the throughput performance of \shoal{} does not yet match that of the underlying Galapagos layer, it can certainly be improved to do so for certain use cases.
The limitations are largely due to incomplete HLS code optimizations for different modules.
After performing these optimizations in the GAScore, \shoal{} should not bottleneck throughput for communication that does not use external memory.
Communication that requires reads and/or writes to external memory will have lower throughput as it results in blocking operations with inherently long and variable access times.

\subsection{Stencil Codes}
\label{sec:eval:jacobi}

Stencil codes represent iterative simulations within a system that is composed of simulation units or cells arranged in a regular matrix.
The state of each cell at $t=0$ represents the initial conditions of the system that then change at discrete time steps.
At $t=n$, the state of each cell can be computed as a function of both the cell's state and that of its neighbors at $t=n-1$.
In this way, the iterative algorithm continues for some defined number of iterations or until a convergence condition is reached.

The Jacobi method is one algorithm in this class where the system is a 2D matrix, which in the simplest case is a square of size $N$.
This algorithm has been tested by prior work in \citeTip{willenberg_heterogeneous_2014} and~\citeTip{pandit_extended_2016} and we adapt it here using \shoal{}.
The von Neumann stencil is used in both works and carried forward here.
This neighborhood only considers the cells in the cardinal directions as neighbors and excludes diagonals.

In software, we adapt the code used in~\cite{pandit_extended_2016} using the \shoal{} API.
While this same code can be annotated with preprocessor directives, run through HLS and implemented on hardware, this approach does not yield good performance.
Instead we divide the software code in two sections: control and computation.
The former includes all \shoal{} API calls and is synthesized to hardware using HLS while the latter is replaced with an optimized VHDL core from \cite{willenberg_heterogeneous_2014}.

\subsubsection{Software Performance}
\label{sec:eval:jacobi:sw}

\begin{figure}[ht]
  \centering
  \includegraphics[width=\linewidth]{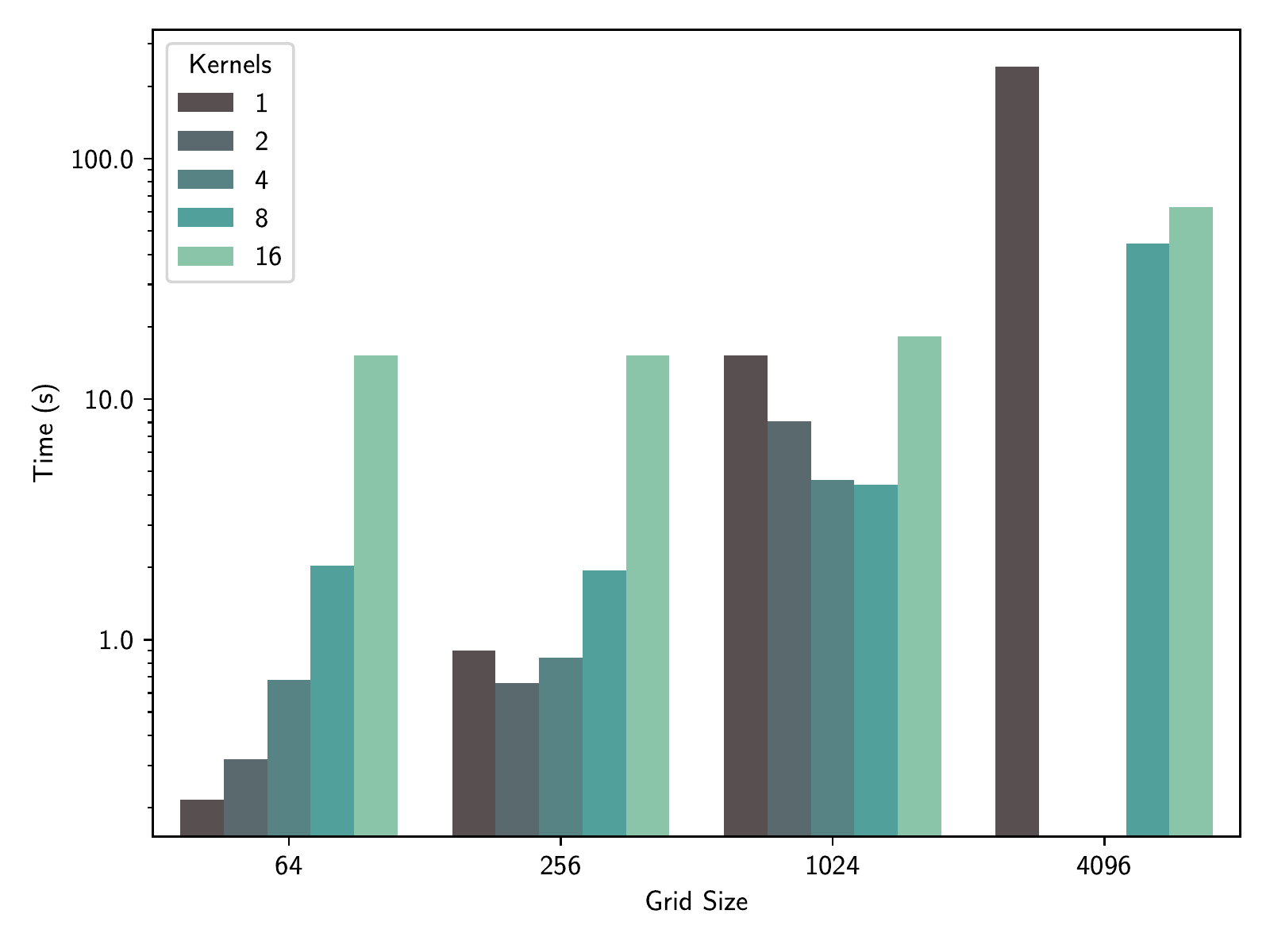}
  \caption{The Jacobi application in software for 1024 iterations}
  \label{img:eval:jacobi:sw}
\end{figure}

Selected results from running the Jacobi application on a single software node are shown in Figure~\ref{img:eval:jacobi:sw} to show a breadth of grid sizes.
The application is run for 1024 iterations and the elapsed time is recorded.
Several interesting trends can be observed here that also apply to the hardware implementations.
For small grid sizes, the overhead of communication, synchronization and memory contention dominates and results in longer execution times as the number of kernels is increased.
At a grid size of 1024, this trend changes and increasing the number of kernels improves the run time to a point.
With 16 kernels on one node, the computation time is almost a second faster than the 8 kernel case but the significantly increased time spent in synchronization offsets this saving.
At the largest tested grid size, we see improvement when increasing the number of kernels again though not with 16 kernels.
We expect the 16-kernel case to be more viable if they are split over more nodes instead of concentrated on one, as we see in the hardware cases.
Note that with a grid size of 4096, using two and four kernels does not currently work.
The grid allocated to each kernel in this case, and therefore the amount of data that must be exchanged at each iteration, is too large to send in a single AM\footnote{Currently, libGalapagos enforces a maximum packet size of 9000 bytes—the size of an Ethernet jumbo frame—due to limitations imposed by the hardware TCP/IP core.}.
The resolution to this limitation is to detect whether the message size exceeds the limit and request the data in smaller sections but this has not been implemented.
With a single kernel, there are no adjacent kernels and so no data is exchanged.

\subsubsection{Hardware Performance}
\label{sec:eval:jacobi:hw}

In the hardware tests, the control kernel remains in software but all computation kernels are moved into one or more FPGAs.
Communication between nodes is performed over TCP to ensure reliability.

Hardware experiments show similar trends to those in software.
For small grid sizes, the cost of communication dominates overall run time.
Until at least a grid size of 2048, it is better to use a single FPGA and a reduced number of kernels.
Having many kernels on a single FPGA creates contention for RAM and decreases performance for these grid sizes.

\begin{figure}[ht]
  \centering
  \includegraphics[width=\linewidth]{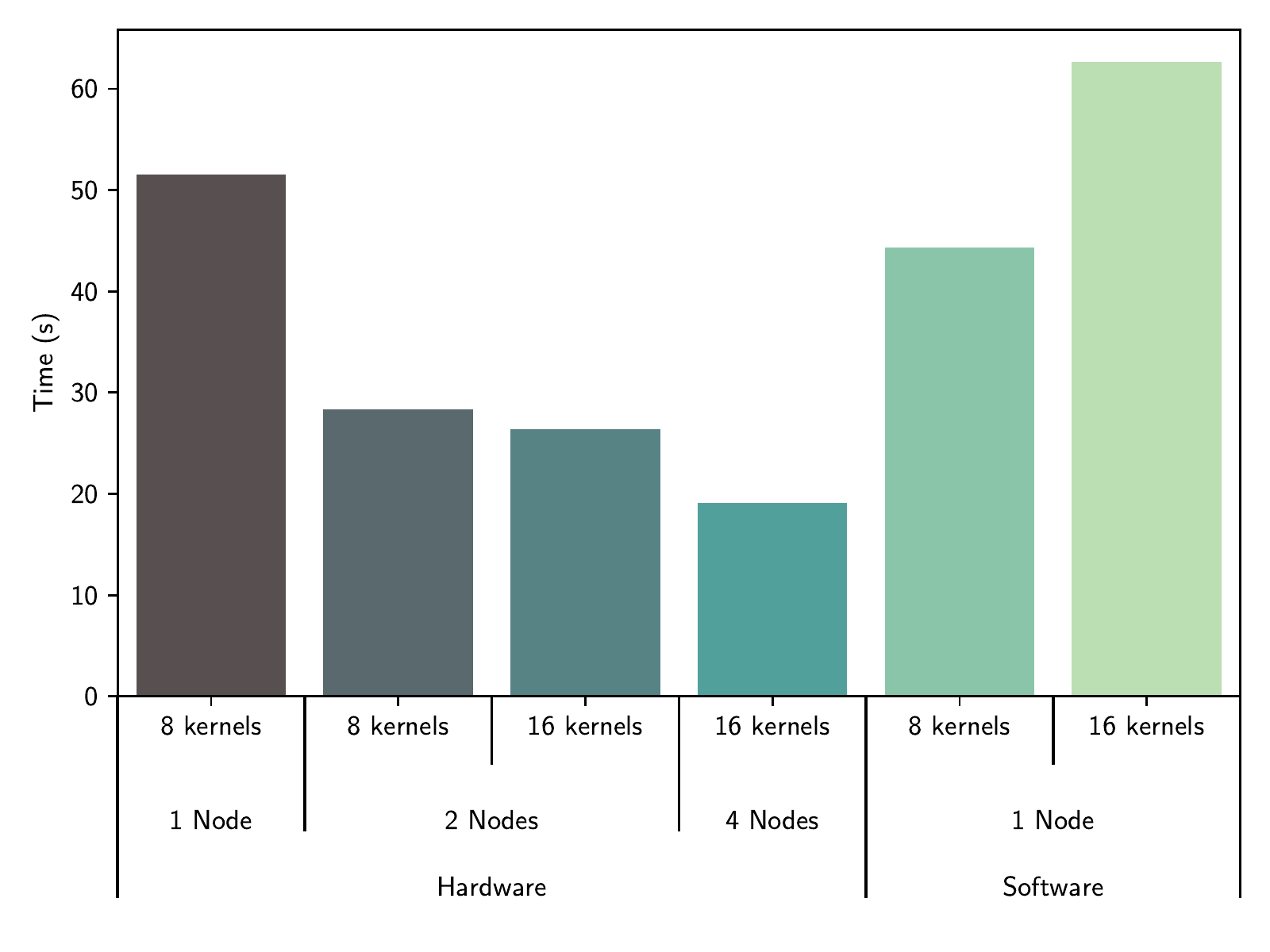}
  \caption[Run time of Jacobi for 1024 iterations at a grid size of 4096 in different topologies]{Run time of Jacobi for 1024 iterations at a grid size of 4096 in different topologies. The bar labels show the total number of computation kernels in the test.}
  \label{img:eval:jacobi:hw}
\end{figure}

We focus here on the larger sizes where kernels and nodes benefit run time.
Figure~\ref{img:eval:jacobi:hw} shows a comparison between hardware and software for the case where the grid size is 4096, iterations are again fixed to 1024 and there are either eight or sixteen kernels in total.
In hardware, holding the total number of kernels constant but spreading them out over multiple nodes improves performance as it decreases contention of local resources.
Increasing the number of kernels also improves run time but not necessarily as dramatically.
With more than one FPGA, the hardware is markedly faster than a single software node.

%% file: tables/utilization_GAScore_1.tex
\begin{table}[ht]
\caption{Hardware utilization of the GAScore (with one kernel) on the 8K5}
\label{table:util:gascore:1}
\centering
\sisetup{
  table-number-alignment = center
}
\begin{tabular}{@{}l
                    S[table-figures-integer = 6, table-figures-decimal = 0, round-mode = places, round-precision = 0]
                    H 
                    S[table-figures-integer = 7, table-figures-decimal = 0, round-mode = places, round-precision = 0]
                    H 
                    S[table-figures-integer = 4, table-figures-decimal = 1, round-integer-to-decimal, round-mode = places, round-precision = 1]
                    H 
                    H
                    H
                    @{}}
\toprule
Component & LUTs & \% & FFs & \% & BRAMs & \% & DSPs  & \% \\
\midrule
\tableindent[0] GAScore & 3595 & 0.5419380125422093 & 4634 & 0.3492824409068982 & 28.0 & 1.2962962962962963 & 0 & 0.0 \\
\tableindent[1] am\_rx & 274 & 7.62 & 377 & 8.1355 & 0.0 & 0.0 & 0 & 0.0 \\
\tableindent[1] am\_tx & 274 & 7.62 & 380 & 8.20 & 0.0 & 0.0 & 0 & 0.0 \\
\tableindent[1] AXI DataMover & 1381 & 38.41 & 1465 & 31.6 & 8.5 & 30.357 & 0 & 0.0 \\
\tableindent[1] FIFOs & 99 & 2.753 & 166 & 3.58 & 2.5 & 8.92857 & 0 & 0.0 \\
\tableindent[1] Interconnects & 600 & 16.6898 & 703 & 15.17 & 0.0 & 0.0 & 0 & 0.0 \\
\tableindent[1] Hold Buffer & 423 & 11.766 & 881 & 19.01 & 8.5 & 30.357 & 0 & 0.0 \\
\tableindent[1] xpams\_rx & 70 & 1.947 & 80 & 1.726 & 0.0 & 0.0 & 0 & 0.0 \\
\tableindent[1] xpams\_tx & 73 & 2.03 & 72 & 1.5537 & 0.0 & 0.0 & 0 & 0.0 \\
\tableindent[1] add\_size & 171 & 4.7566 & 157 & 3.388 & 8.5 & 30.357 & 0 & 0.0 \\
\tableindent[1] Handler Wrapper & 229 & 6.3699582753824755 & 353 & 7.6176089771255935 & 0.0 & 0.0 & 0 & 0.0 \\
\tableindent[2] Handler 0 & 228 & 99.56331877729258 & 345 & 97.73371104815864 & 0.0 & 0.0 & 0 & 0.0 \\
\midrule
Alpha Data 8K5 & 663360 & 0 & 1326720 & 0 & 2160 & 0 & 5520 & 0 \\
\bottomrule
\end{tabular}
\end{table}

%% file: sections/conclusion.tex
\section{Conclusions}
\label{ch:conclusion}

The rise of hardware-as-a-service in the cloud and strategic acceleration of data center applications with FPGAs motivates the need to develop applications that can take advantage of a heterogeneous cluster.
Through HLS, the problem can be opened up to the numerous domain experts who have potential applications but lack the hardware skills to implement their vision.
Note that this is not an endorsement of current HLS solutions.
Writing performant HLS code is difficult and the tool may implement something else entirely.
However, it can be a valuable first step to explore the design space and implement the scaffolding in which hardware experts can insert optimizations.

Writing applications in a distributed environment requires the use of a robust communication API, which is the main goal of this work.
The previous research in THeGASNet and THe\_GASNet projects had to incorporate their own custom and platform-specific communications infrastructure in addition to the PGAS functionality.
They demonstrated important concepts but were not portable.
Leveraging that previous work, we showed how to implement PGAS functionality as a layer on top of the Galapagos platform.
Our work then becomes much more portable and can be used wherever a Galapagos platform can be implemented.
We show that this PGAS layer requires a very low resource overhead and adds little to the underlying Galapagos latency.

We have developed a cross-platform API that provides a mix of message types.
We demonstrate this library using a Jacobi method application to highlight the ease of use in adding FPGAs in a high-compute task.
Instead of allocating time to setup the network and access to memory on the FPGA, time can be devoted to developing the application that can be run on any topology and platform with few or no changes in source code.

\subsection{Future Work}
\label{sec:conclusions:future}

In addition to optimizations to improve \shoal{}, another avenue of future work is to change the \shoal{} API to be more modular by defining discrete components that can be selectively enabled on a particular \shoal{} application.
\shoal{} is currently implemented as a monolith based on the specification in THeGASNets.
As such, \shoal{} must be fully flexible to handle all message types.
However, it is likely that a given application only uses a subset of the specification, resulting in constant cost to evaluate conditions that will never be true and unnecessary hardware usage on the FPGA.
With a modular API specification, we can define discrete components of the API that can be selectively enabled.
The freedom to choose can also be used to implement different communication models.
For example, enabling barriers and Medium messages only creates a simple point-to-point communication protocol that can be used as a thin layer on top of libGalapagos or used to implement message passing.